\documentclass[a4paper,11pt]{article}

\usepackage[left=2.5cm,right=2.5cm,top=3cm,bottom=3cm]{geometry}

\usepackage{amsfonts,amsmath,mathtools,mathrsfs,amssymb,amsthm}
\usepackage{hyperref}
\usepackage[noadjust]{cite}
\usepackage{bm}
\usepackage{dsfont}
\usepackage{graphicx}
\usepackage{xcolor}
\usepackage{float}
\usepackage{braket}
\usepackage{graphicx}
\usepackage{caption}
\allowdisplaybreaks
\bibliography{references}
\usepackage[font=small,labelfont=bf]{caption}

\newtheorem{theorem}{Theorem}
\newtheorem{remark}[theorem]{Remark}
\newcommand{\R}{\mathbb{R}}
\newcommand{\Z}{\mathbb{Z}}
\newcommand{\Hi}{\mathcal{H}}
\newcommand{\ketbra}[2]{|#1\rangle\langle#2|}
\newcommand{\abs}[1]{\left|#1\right|}
\newcommand{\de}{\mathrm{d}}
\newcommand{\w}{\bm{w}}
\newcommand{\x}{\bm{x}}
\newcommand{\sgn}{\mathrm{sgn}}
\newcommand{\inpset}{\mathscr{I}}
\newcommand{\labset}{\mathscr{X}}
\newcommand{\tr}{\mathrm{tr}}

\newcommand{\ini}{\mathrm{in}}
\newcommand{\s}{\sigma}

\newcommand{\out}{\mathrm{out}}
\newcommand{\rme}{\mathrm{e}}
\newcommand{\ii}{\mathrm{i}}
\newcommand{\cm}{``}

\begin{document}

\title{Pseudo quantum advantages in perceptron storage capacity}

\author{Fabio Benatti$\hspace{0.4mm}^{1,2}$\thanks{Email: benatti@ts.infn.it}\;, \hspace{1.2mm}
Masoud Gharahi$\hspace{0.4mm}^{1}$\thanks{Email: masoud.gharahi@gmail.com (Present address: \textit{Faculty of Physics, Astronomy and Applied Computer Science, Jagiellonian University, 30-348 Kraków, Poland})}\;,
\hspace{1.2mm}
Giovanni Gramegna$\hspace{0.4mm}^{3,4}$\thanks{Email: giovanni.gramegna@uniba.it}\;,
\\
Stefano Mancini$\hspace{0.4mm}^{5,6}$\thanks{Email: stefano.mancini@unicam.it}\;, \hspace{0.6mm}
and Vincenzo Parisi$\hspace{0.4mm}^{5,6}$\thanks{Corresponding author; email: vincenzo.parisi@unicam.it. (Present address: \textit{CONCEPT Lab, Fondazione Istituto Italiano di~Tecnologia,
via E.~Melen 83, Genova, 16152, Italy})}}

\date{
{\normalsize
$^1$\textit{Department of Physics, University of Trieste, Strada Costiera 11, I-34151, Trieste, Italy}\\	
\vspace{2mm}
$^2$\textit{Istituto Nazionale di Fisica Nucleare, Sezione di Trieste, Strada Costiera 11, I-34151,\\
Trieste, Italy}\\
\vspace{2mm}
$^3$\textit{Dipartimento di Fisica, Universit\`a degli Studi di Bari, I-70126 Bari, Italy}\\
\vspace{2mm}
$^4$\textit{Istituto Nazionale di Fisica Nucleare, Sezione di Bari, I-70126,\\
Bari, Italy
}\\
\vspace{2mm}
$^5$\textit{School of Science and Technology, University of Camerino,\\
Via Madonna delle Carceri, 9, Camerino, I-62032, Italy}\\
\vspace{2mm}
$^6$\textit{Istituto Nazionale di Fisica Nucleare, Sezione di Perugia,\\
via A.~Pascoli, I-06123 Perugia, Italy}
}	
}
	
\maketitle
	
\begin{abstract}
We investigate a generalized quantum perceptron architecture characterized by an oscillating activation function with a tunable frequency ranging from zero to infinity. Employing analytical techniques from statistical mechanics, we derive the optimal storage capacity and demonstrate that the classical result is recovered in the limit of vanishing frequency. As the frequency increases, however, the architecture exhibits enhanced quantum storage capabilities. Notably, this improvement stems solely from the specific form of the activation function and, in principle, could be emulated within a classical framework. Accordingly, we refer to this enhancement as a \emph{pseudo quantum advantage}.
\end{abstract}

\tableofcontents


\section{Introduction and motivation}

Recent advancements in quantum computing have enabled the implementation of machine learning concepts on quantum hardware~\cite{biamonte2017}. This development raises the prospect of quantum neural networks outperforming their classical counterparts, offering enhanced storage capacity and superior information processing capabilities.
In the classical framework, using statistical mechanical tools, deep connections among neural networks, spin glasses, and information processing have been
uncovered~\cite{enge2000,nish2001}. A significant advantage of the statistical mechanics approach is its ability to extract global, macroscopic features of physical systems without requiring detailed knowledge of their microscopic details. This methodology has also enabled the characterization of artificial neural networks without requiring prior knowledge of specific learning rules. Instead, it approaches the problem by treating network weights as random variables, a framework often referred to as \emph{Gardner’s program}~\cite{gard1987,gard1988,gard1988s}.

In the quantum setting, the statistical approach has already been used to estimate the storage capacity of  continuous and discrete versions of quantum perceptrons; namely, of the fundamental building blocks of quantum neural networks.
Yet, the models that have been considered show apparently opposite results regarding their storage capabilities.
Indeed, the results in~\cite{bena2022,bena2024} indicate that quantum advantages in storage properties are unlikely. There, a natural quantum
encoding of the classical patterns and a binary classification rule of the measurement outcomes are provided which yield an optimal storage capacity always bounded from above by the maximal classical storage capacity of $\alpha_c=2$.
The reason behind such a behavior  is due to the fuzziness injected into the procedure by the non-perfect distinguishability of the quantum states encoding the classical patterns and by the intrinsic randomness of the measurement results
upon which the classification of the quantumly encoded classical patterns is operated.

Conversely, the results in~\cite{gratsea2024}
suggest that the quantum storage capacity may become twice
its classical counterpart, yet
this result
stems from a sign ambiguity inherent to the measurement process in the employed perceptron model.  Furthermore, in~\cite{urushi2024}, the authors investigated a quantum perceptron implemented on a quantum circuit using a repeat until success method, finding a storage capacity almost five times larger than the classical one.
Again, this advantage ought to
be ascribed to the highly non-linear form of the activation function resulting from the employed quantum perceptron model.

In this work, taking inspiration from the latter analysis, we apply Gardner’s program to
a broadly defined quantum perceptron architecture~\cite{torr2019} featuring an oscillating activation function whose frequency can range from zero to infinity. We compute analytically the

storage
capacity and show that, while the classical result is recovered at vanishing frequency, increasing the frequency yields enhanced --- and even diverging
--- quantum storage capacity. However, since this effect arises solely from the form of the activation function, it can, in principle, be replicated within a classical framework. Therefore, we refer to this enhancement in storage capacity as a \emph{pseudo quantum advantage}.


\section{Basic tools}
This section deals with the main tools and techniques necessary for our later purposes. We start by recalling the main features of a so-called \emph{classical perceptron}~\cite{ros1957,ros1958,mccu1943,good2016,palm2018}. Then, we introduce the notion of storage capacity as an appropriate parameter to characterize the perceptron performances. Finally, we outline the essential features of Gardner’s approach to evaluate the storage capacity~\cite{gard1987,gard1988,gard1988s,enge2000,nish2001}.


\subsection{The classical perceptron}
Artificial neurons are the fundamental building blocks of an artificial neural network. From the mathematical point of view, the output of an artificial neuron can be modeled by a map
\begin{equation}\label{eq.1}
\mathbb{R}^N\ni\x\mapsto f\big(\w\cdot\x+b\big)\in\mathbb{R},
\end{equation}
where $\x=(x_1,\ldots,x_N)$, $\w=(w_1,\ldots,w_N)$ represent the \emph{input patterns} and the \emph{vector of weights}, respectively,
while $\w\cdot\x=\sum_{i=1}^N w_ix_i$. The constant term $b\in\mathbb{R}$ is the so-called \emph{bias}, while $f\colon\mathbb{R}\rightarrow \mathbb{R}$ is a non-linear function --- i.e., the \emph{activation function} --- which determines whether the artificial neuron is active or not. In the case where the activation function is chosen to be the Heaviside theta function --- that is,  $f(x)=\Theta(x)=1$ if $x>0$, and zero otherwise --- or the sign function, $\sgn(x)$, the artificial neuron is called \emph{perceptron}~\cite{ros1957,ros1958,mccu1943,good2016,palm2018}.

Let $f(x)\equiv\sgn(x)$ and, for simplicity, set the bias $b=0$ as in the
standard task addressed by a classical perceptron. The latter is the so-called binary classification problem, consisting in the assignment of a given input vector to one of two possible classes~\cite{good2016,palm2018} specified by the values of a binary variable $\xi=\pm1$. More precisely, let us introduce the following sets
\begin{equation}
\inpset\coloneqq\{\x^\mu\in\{-1,1\}^N\mid 1\leq \mu\leq p\},\quad \labset\coloneq\{\bm{\xi}=\{\xi^\mu\}_{\mu=1}^p\mid\xi^\mu=\pm 1\},
\end{equation}
which represent the input set --- here we are considering input vectors with binary entries --- and the label set, respectively. Depending on the weight vectors $\w$ in $\mathbb{R}^N$, a classical perceptron provides a classification of $p$ input patterns $\{\x^\mu\}_{\mu=1}^p\subset\inpset$. Given an assigned classification vector $\bm{\xi}\in\labset$ to be implemented, with the chosen activation function the perceptron outputs $\sgn(\w\cdot\x^\mu)$, so that the classical perceptron correctly classifies the input patterns iff
\begin{equation}\label{eq.4}
\sgn(\w\cdot\x^\mu)=\xi^\mu,\quad\forall\mu=1,\ldots,p.
\end{equation}
Condition~\eqref{eq.4} is sometimes rephrased in terms of the so-called \emph{pattern stabilities}, defined as
\begin{equation}
\Delta^\mu\coloneqq\xi^\mu\,\w\cdot\x^\mu\ .
\end{equation}
Accordingly, the classification of the input pattern is correct --- w.r.t.\ a given chosen target classification $\bm{\xi}$ --- iff the condition
\begin{equation}\label{eq.6}
    \Delta^\mu\geq 0
\end{equation}
holds for any $\mu=1,\ldots,p$. In many practical situations, the above condition~\eqref{eq.6} is strengthened by requiring $\Delta^\mu>\kappa>0$ for all $\mu=1,\ldots,p$.
Indeed, this condition provides a higher stability of the perceptron, as it prevents incorrect classification due to noise in the input pattern components.


\subsection{The storage capacity problem}\label{subsec.2.2}

A paramount feature of an artificial neuron is its ability to store and classify input patterns. This performance is typically assessed by the so-called \emph{storage capacity}; namely, by a threshold parameter associated with the volume of input patterns a classical perceptron can correctly classify when their
dimension $N$ increases~\cite{enge2000,nish2001}.

A convenient statistical way to define and compute the storage capacity of a classical perceptron is by means of Gardner's approach~\cite{gard1987,gard1988,gard1988s}; here, one starts by introducing the normalized volume --- i.e., the Gardner volume --- defined as
\begin{equation}\label{eq.8}
V_N\coloneqq\frac{1}{Z_N}\int\de\mu(\w)\prod_{\mu=1}^p\Theta(\Delta^\mu-\kappa),
\end{equation}
where $\Theta$ is the Heaviside function, $\de\mu(\bm{w})$ is the normalized  uniform measure on the $N$-dimensional sphere of radius $\sqrt{N}$:
\begin{equation}\label{ZN}
\de\mu(\bm{w})=\frac{1}{C_N}\int\de \w\,\delta\big(\|\w\|^2-N\big) , \quad C_N=\int\de \w\,\delta\big(\|\w\|^2-N\big)=\frac{2\pi^{N/2}N^{(N-1)/2}}{ \Gamma (N/2)} ,
\end{equation}
where $\de\w=\prod_{i=1}^N\de w_i$\footnote{Note that, the measure defined as $\lambda(E)\coloneqq\int_E\de \w \delta(\|\w\|^2-N)\prod_{\mu=1}^p\Theta(\Delta^\mu-\kappa)$, for every Borel subset $E$ of $\mathbb{R}^N$, is a well defined Radon measure on $\mathbb{R}^N$.}, and $ \Gamma (z)$ is the Euler Gamma function. The Gardner volume \(V_N\) quantifies the fraction of weight vectors
\(\w\in\mathbb{R}^N\) constrained to the sphere of radius \(\sqrt{N}\)
that correctly classify a set of \(p\) input patterns. In the thermodynamic
limit \(N\to\infty\), one is interested in the regime where a macroscopic number
of patterns can be stored, i.e.\ \(p=\alpha N\) with fixed load \(\alpha>0\).
As \(N\) increases, the \(p=\alpha N\) classification constraints in~\eqref{eq.8} typically reduce \(V_N\) at least exponentially in \(N\).
The problem becomes unfeasible when this reduction is super-exponential,
in which case \(V_N\) vanishes too rapidly to allow storage. The storage
capacity is thus defined as the critical value \(\alpha_c\) of the load
parameter \(\alpha=p/N\) that separates the regime where \(V_N\) remains
exponentially small in \(N\) from the regime where it decays
super-exponentially~\cite{enge2000,nish2001}.

To compute the storage capacity, one can exploit the formal analogy between  { expression~\eqref{eq.8} and a statistical mechanics partition function.
Then, we will consider the patterns $\x^\mu$ and the labels $\bm{\xi}^\mu$
as independent random variables with independent and identically distributed  entries, according to}
\begin{equation}\label{eq:distPatternsLabels}
P(x^{\mu}_j=1)= P(x^{\mu}_j=-1)=\frac{1}{2},\qquad P(\xi^\mu=1)=P(\xi^\mu=-1)=\frac{1}{2}.
\end{equation}
As a consequence, the volume~\eqref{eq.8} becomes a random variable whose typical value $V_N\sim \rme^{N\mathcal{F}}$ is characterized by the free energy~\cite{meza1987,Malatesta}
\begin{equation}\label{eq:FreeEn}
    \mathcal{F}(\alpha)=\lim_{\substack{p,N\rightarrow \infty\\p/N=\alpha}}\frac{\langle\ln V_N\rangle_{\inpset,\labset}}{N}.
\end{equation}
In order to compute the so called quenched average $\langle \ln V_N \rangle_{\inpset,\labset}$, a notoriously difficult task, a most convenient approach is the so-called \emph{replica trick}~\cite{meza1987,tala2011}.
\begin{equation}\label{eq:replicaTrick}
\langle \ln V_N\rangle_{\inpset,\labset}=\lim_{n\rightarrow 0}\frac{\langle V_N^n\rangle_{\inpset,\labset} -1}{n}=\lim_{n\rightarrow 0}\frac{\ln\langle V_N^n\rangle_{\inpset,\labset}}{n}\ ,
\end{equation}
but evaluating $\langle V_N^n\rangle_{\inpset,\labset}$ for $n$ integer.

The replica trick~\eqref{eq:replicaTrick} allows us to express the expectation value $\langle \ln V_N\rangle_{\inpset,\labset}$ in terms of the average volume over $n$  replicas of the single perceptron random setting
\begin{equation}\label{eq.11}
\langle V_N^n\rangle_{\inpset,\labset}=\int\prod_{ \gamma =1}^n\de\mu(\w_\gamma )\left\langle\prod_{\mu=1}^p\prod_{ \gamma =1}^n \Theta\big(\xi^\mu\,\w_\gamma \cdot\x^\mu-\kappa\big)\right\rangle_{\inpset,\labset},
\end{equation}
which can be computed in the $N\rightarrow \infty$ limit through a saddle-point approximation. In the replica symmetric scenario, the computation is performed with the introduction of the order parameter $q$, whose value at the saddle point characterizes the typical overlap between two different replicas $\w_1,\w_2$ extracted from the uniform measure within the solution space~\cite{Watkin,Malatesta}:
\begin{equation}
    q=\left\langle\int\de\mu(\w_1)\de\mu(\w_2) \frac{\w_1\cdot\w_2}{N}\prod_{\mu=1}^p \Theta(\xi^\mu \w_1\cdot \x^\mu-\kappa)\Theta(\xi^\mu \w_2\cdot \x^\mu-\kappa) \right\rangle.
\end{equation}
The value of $q$ at the saddle point depends on the value of the load parameter $\alpha$ in~\eqref{eq:FreeEn}, and in particular one finds that it monotonically increases in $\alpha$. Then, the critical value $\alpha_c$ can be characterized as the value of $\alpha$ such that $q\to 1$, signaling the fact that the typical volume of the solution space shrinks to zero.
In the classical perceptron, this procedure yields the final result for the storage capacity in the form~\cite{gard1987, gard1988,palm2018}
\begin{equation}
\alpha_c(\kappa) =
\left(
\int_{-\kappa}^{+\infty}
\frac{{\rm d}y}{\sqrt{2\pi}}\,
{\rm e}^{-y^2/2} (\kappa + y)^2
\right)^{-1}.
\label{eq:alphac}
\end{equation}
Interestingly, when $\kappa\rightarrow 0$, one finds $\alpha_c(0)=2$, which is the value of the storage capacity obtained through a completely different approach, based on a geometric argument due to Cover~\cite{cover1965}.


\section{Quantum storage capacity}

Quantum neural networks generalize at the quantum level the notion of a feed-forward neural network. Like their classical counterparts, the fundamental computational unit of a quantum neural network is represented by a so-called quantum perceptron. In recent years, several proposals for implementing a quantum perceptron have been considered~\cite{bena2019,bena2022,bena2024,toth1996,Mccl2016,rome2021,mita2018,mccl2018,bene2019,gupt2001,andr2002,zhou2012,dasi2016,wan2017,schu2021,torr2019,cao2017}. Regardless of the specific model, each approach typically involves three essential steps: first, constructing an encoding circuit to map classical input data into a quantum state; second, defining a set of trainable quantum gates --- controlled by  tunable weights $w_{ij}$ --- to realize a quantum counterpart to the non-linear output of the classical perceptron; and finally, setting a readout operation to retrieve a classical output from the quantum system. In this respect, notice that the quantum measurement process itself, in its selective version, amounts to a non-linear operation.


\subsection{A discrete model of quantum perceptron}
\label{sec:2.1}
In this work, we focus on a model first proposed in~\cite{torr2019}. A perceptron is here implemented by a qubit subjected to an external coupling that operates a unitary transformation parametrized by a classical activation function. Specifically, the $k$-th qubit in a multi-layered perceptron is acted upon  by a unitary transformation $U_k(\w_k,f)$ that depends on a non-linear activation function $f$ and on tunable weights $w_{kj}$, $j=1,2,\ldots,k-1$, as follows:
\begin{equation}\label{eq.24}
U_k(\w_k,f)\coloneqq\exp\bigg(-\ii \arcsin\sqrt{f\Big(\mbox{$\sum_{j<k}$}w_{kj}\sigma_z^{(j)}-b_k\Big)}\otimes\sigma_y^{(k)}\bigg),
\end{equation}
where $\sigma_\alpha^{(i)}$, $\alpha\in\{x,y,z\}$, denotes the $\alpha$-Pauli operator acting on the $i$-th qubit.

\begin{remark}\label{rem2}
The above model of discrete quantum perceptron has been proved to contain the classical perceptron as a limit and thus to  provide a universal approximator of continuous functions~\cite{torr2019}.
\end{remark}

In the simplest case, the quantum neural network consists of a single input layer with $N$ qubits, and an output layer with a single qubit. Let $\Hi_N=\mathbb{C}^{2^N}$ and $\ketbra{\Psi_N}{\Psi_N}$ denote the Hilbert space, and a pure state of the input layer, respectively. Similarly, let $\Hi_\out=\mathbb{C}^2$ and $\ketbra{\phi}{\phi}$ be the Hilbert space and an initial pure state of the output quantum neuron.
Without loss of generality, we can choose
the
initial state of the network in a factorized form, i.e.,
\begin{equation}\label{initialstate}
\rho_\ini\coloneqq\ketbra{\Psi_N}{\Psi_N}\otimes\ketbra{\phi}{\phi}\ .
\end{equation}
Then,
setting the bias $b=0$ and
letting $f$ be the Heaviside $\Theta$ function, the unitary action~\eqref{eq.24} implementing the quantum perceptron reduces to
\begin{equation}
\label{unitarytheta}
U(\w,\Theta)=\exp\left(-\ii \arcsin\sqrt{\Theta(\w\cdot\bm{\s}_z)}\otimes\sigma_y^{(\out)}\right),
\end{equation}
where $\bm{\s}_z=(\s_z^{(1)},\sigma_z^{(2)}\ldots,\sigma_z^{(N)})$,  $\w=(w_1,\ldots,w_N)\in\mathbb{R}^N$, and $\sigma_y^{(\out)}$ is the $y$-Pauli operation performed by the output perceptron.
An input vector of binary classical data $\x=(x_1,x_2,\ldots,x_N)\in\{\pm\}^N$ is naturally encoded as the tensor product of eigen-states of $\sigma_z$; namely,
 \begin{equation}\label{eq.33n}
 \{\pm\}^N\ni\bm{x}\mapsto \ket{\bm{x}}=\ket{x_1}\otimes\ket{x_2}\otimes\ldots\otimes\ket{x_N},\quad \sigma_z^{(j)}\ket{x_j}=x_j\ket{x_j}.
 \end{equation}
With this notation, the unitary in~\eqref{unitarytheta} reads
\begin{align}\nonumber
U(\w,\Theta) &= \sum_{{\bf{y}}\in\{\pm\}^N}\ket{\bf{y}}\bra{\bf{y}}\otimes \exp\Big(-\ii \arcsin\left (\sqrt{\Theta(\w\cdot\bf{y})}\right)\,\s_y^{(\out)}\Big) \\ \label{eq.28}
&=
\sum_{{\bf{y}}\in\{\pm\}^N}\ket{\bf{y}}\bra{\bf{y}}\otimes\Big(\sqrt{1-\Theta(\w\cdot\bf{y})}\mathbb{I}^{(\out)}-\ii \sqrt{\Theta(\w\cdot{\bf{y}})}\,\s_y^{(\out)}\Big).
\end{align}
By choosing the initial state in~\eqref{initialstate} as the projector onto $\ket{\x}\otimes\ket{-1}$, where
$\sigma^{\out}_z\ket{-1}=-\ket{-1}$, the state of the output qubit becomes
\begin{align}\label{eq.28s}
\rho_{\w,\x}^{(\out)} &= \tr_{\Hi_N}\Big(U(\w,\Theta)\big(\ketbra{\x}{\x}\otimes\ket{-1}\bra{-1}\big)U^\dagger(\w,\Theta)\Big)\\
\label{eq.36s}
&= \big(1-\Theta(\w\cdot\x)\big)\ketbra{-1}{-1}+
\Theta(\w\cdot\x)\ketbra{1}{1}.
\end{align}
The last readout step of the perceptron architecture evaluates the expectation value of $\sigma_z$ w.r.t.\
the output state $\rho^{(\out)}_{\w,\x}$, yielding
\begin{equation}
\label{meanvalue}
\langle\sigma_z\rangle_{\w,\x}=2\,\Theta(\w\cdot\x)-1.
\end{equation}

\begin{remark}
\label{remConv}
{The choice of the operator to measure in the readout step is completely arbitrary and should be chosen out of convenience. A measurement in a different basis can always be thought of as an additional fixed unitary step (independent on the data to be classified or the trainable parameter $\w$) just before the measurement is performed.}
\end{remark}

\begin{remark}\label{rem3}
The discrete quantum perceptron model outlined above is closely related to another proposal put forward in~\cite{cao2017} whereby a unitary gate
\begin{equation}
U(z)=\exp\big(\ii \beta^{(j)}(z)\sigma_y\big)\ ,\qquad
\beta^{(j)}(z)=2\arctan\big(\tan^{2^j}(z)\big),
\end{equation}
is implemented by a so-called \cm repeat until success" strategy.
Namely,
$U(z)$ is a rotation around $y$ by an angle $\beta^{(j)}(z)$ that, in the limit for $j\rightarrow \infty$, converges to a step-wise function in the interval $z\in\big[-\frac{\pi}{4},\frac{\pi}{4}\big]$.
\end{remark}


\subsection{Quantum storage capacity}\label{subsec.3.1}

In order to compute the storage capacity of the quantum perceptron presented in Section~\ref{sec:2.1}, we follow Gardner's approach as discussed in Section~\ref{subsec.2.2}.
We focus upon the volume~\eqref{eq.8} of those (normalized) weight vectors $\bm{w}\in\R^N$ which correctly classify $p$ input patterns $\x^\mu\in\{\pm\}^N$, i.e.,
\begin{align}\label{eq.40}
V(\{\x^\mu,\xi^\mu\}_{\mu=1}^p)&=\int \de \mu(\bm{w})\prod_{\mu=1}^{p}\Theta(\xi^\mu\langle\sigma_z\rangle_{\w,\x})\nonumber\\
&=\int \de \mu(\bm{w})\prod_{\mu=1}^p \Theta\Big(\xi^\mu\big(2\,\Theta(\w\cdot\x^\mu)-1\big)\Big),
\end{align}
where, for simplicity, we set the stabilizing threshold $\kappa=0$. By observing that
\begin{equation}
\Theta\Big(\xi^\mu\big(2\,\Theta\big(\w\cdot\x^\mu\big)-1\big)\Big)=\Theta\big(\xi^\mu\w\cdot\x^\mu\big),
\end{equation}
i.e., the volume~\eqref{eq.40} is formally equivalent to the Gardner volume~\eqref{eq.8} of the classical perceptron, we get to the conclusion that it will yield the same limiting value $\alpha_c(0)=2$.

The above result suggests that, in the quantum perceptron model described in Section~\ref{sec:2.1},
no quantum advantage can be observed at the level of the storage capacity.
Notice that the unitary operation in~\eqref{unitarytheta} contains the classical non-linear activation function $\Theta(\w\cdot\x)$ depending on the rotation angle. However, a non-linear quantum operation is built in the
quantum circuit through the selective measurement processes on the output $\sigma_z$ that extract the eigenvalues
$\pm 1$ with which the mean-value
$\langle\sigma_z\rangle_{\w,\x}$ is constructed.
We will base our subsequent considerations on the quantum non-linear effects inscribed in the above model (due to the quantum measurement), while
eliminating the classical non-linearity (due to the $\arcsin$ function). We thus modify
the unitary gate of the model in~\cite{torr2019} into
\begin{equation}\label{eq.41s}
    U(\w,\lambda)\coloneq\exp\Big(-\frac{\ii}{2}\frac{\lambda}{\|\bm{w}\|}\w\cdot\bm{\s}_z\,
 \otimes\,\s_y^{(\out)}\Big).
\end{equation}
In the above unitary operator we have also included a \emph{modulation parameter} $\lambda\in[0,+\infty)$ which will allow us to vary the frequency of the oscillations. The removal of non-linearity in the controlled phase of the unitary \eqref{eq.41s} has the advantage of simplifying the practical implementation  with respect to~\eqref{unitarytheta}. As a matter of fact, the unitary \eqref{eq.41s} can be even easily implemented on any NISQ device, requiring only two qubits.

When acting on the multiple eigenstates $\ket{\x}$ in~\eqref{eq.33n}, the unitary operator $U(\w,\lambda)$ provides a rotation around $y$ by an angle $\w\cdot\x$ yielding the output readout state (see~\eqref{eq.28s})
\begin{align}
\label{casestudy}
\hskip-.3cm \rho^{(\out)}_{\w,\x}=&\cos^2\Big(\lambda\frac{\w\cdot\x}{2\|\w\|}\Big)\ketbra{-1}{-1}+\sin^2\Big(\lambda\frac{\w\cdot\x}{2\|\w\|}\Big)\ketbra{1}{1}
+\frac{1}{2}\,\sin\Big(\frac{\lambda \w\cdot\x}{\|\w\|}\Big)\,\big(\ketbra{-1}{1}+\ketbra{1}{-1}\big).
\end{align}
We will consider the measurement of $\sigma_x$ on the output, whose expectation value on~\eqref{casestudy} reads
\begin{equation}\label{eq:sinActivation}
\langle\sigma_x\rangle_{\w,\x}=\sin\Big(\frac{\lambda \w\cdot\x}{\|\w\|}\Big).
\end{equation}
Compared with~\eqref{meanvalue} we can realize that this outcome corresponds to a (sinusoidally)  oscillating activation function with frequency controlled by $\lambda$.
The choice to measure $\sigma_x$ rather than $\sigma_z$ is just a matter of convenience (see Remark~\ref{remConv}): For input data drawn from the~\eqref{eq:distPatternsLabels}, and spherical weights $\w$, the argument of the $\sin$ in~\eqref{eq:sinActivation} is centered around zero, which allows one to easily match the desired distribution of the target classifications in~\eqref{eq:distPatternsLabels}. Moreover, with this choice the classical limit just corresponds to a linearization of the $\sin$, obtained here through $\lambda\rightarrow0$. Exchanging $\sigma_x$ with $\sigma_z$ in this model would just correspond to the introduction of a fixed bias in the model, with unnecessary complications.

Inserting~\eqref{eq:sinActivation} into~\eqref{eq.8}, provides the following Gardner volume
\begin{equation}\label{eq.44s}
V^\lambda_N(\{\x^\mu,\xi^\mu\}_{\mu=1}^p)
=\int\de\mu(\bm{w})\prod_{\mu=1}^p \Theta\Big(\xi^\mu\sin\Big(\frac{\lambda \w\cdot\x^\mu}{\|\w\|}\Big)\Big)\ .
\end{equation}
Following the standard procedure outlined in section \ref{subsec.2.2} we find the free energy in the replica symmetric ansatz (see Appendix~\ref{app.1} for the details)
\begin{equation}\label{eq:FreeEns}
    \mathcal{F}(\lambda,\alpha)=\lim_{\substack{p,N\rightarrow \infty\\p/N=\alpha}}\frac{\langle\ln V_N^\lambda\rangle_{\inpset,\labset}}{N}=\lim_{\substack{p,N\rightarrow \infty\\p/N=\alpha}}\lim_{n\rightarrow 0}\frac{\ln\langle (V_N^\lambda)^n\rangle_{\inpset,\labset}}{nN}=\underset{0\leq q<1}{\mathrm{extr}}\ \mathcal{G}(\lambda,\alpha,q),
\end{equation}
where
\begin{equation}
    \mathcal{G}(\lambda,\alpha,q)=\alpha \int_\mathbb{R} D\omega\ln\Psi(\lambda,q,\omega)+\frac{1}{2}\left[\frac{q}{1-q}+\ln(1-q)\right].
\end{equation}
The saddle point equation $\partial_q\mathcal{G}(\lambda,\alpha,q)=0$ then yields a relation between the load parameter $\alpha$ and the typical overlap $q$:
\begin{align}
\alpha(\lambda,q)&=-\frac{q}{2(1-q)^2}\bigg(\int\frac{\de \omega}{\sqrt{2\pi}}\,{\rm e}^{-\frac{\omega^2}{2}}\partial_q\ln\Psi(\lambda,q,\omega)\bigg)^{-1}\nonumber\\[0.2em]
&=-\frac{q}{2(1-q)^2}\bigg(\int\frac{\de \omega}{\sqrt{2\pi}}\,{\rm e}^{-\frac{\omega^2}{2}}\frac{\Phi(\lambda,q,\omega)}{\Psi(\lambda,q,\omega)}\bigg)^{-1},
\end{align}
where we set
\begin{equation}\label{eq.46s}
\Psi(\lambda,q,\omega)\coloneqq\sum_{k\in\Z}\int_{\epsilon_1(\lambda,q,\omega,k)}^{\epsilon_2(\lambda,q,\omega,k)}\frac{\de z }{\sqrt{2\pi}}\,{\rm e}^{-\frac{ z ^2}{2}},\quad \Phi(\lambda,q,\omega)\coloneqq \partial_q\Psi(\lambda,q,\omega),
\end{equation}
with $\epsilon_1(\lambda,q,\omega,k)$ and $\epsilon_2(\lambda,q,\omega,k)$ given by
\begin{equation}
\epsilon_1(\lambda,q,\omega,k)\coloneqq \frac{-\lambda\sqrt{q}\omega+2k\pi}{\lambda\sqrt{1-q}},\qquad\epsilon_2(\lambda,q,\omega,k)\coloneqq
\frac{-\lambda\sqrt{q}\omega+(2k+1)\pi}{\lambda\sqrt{1-q}}\ .
\end{equation}
The critical storage capacity $\alpha_c(\lambda)$ is then obtained via the limit
\begin{equation}\label{eq.45n}
\alpha_c(\lambda)
=\lim_{q\rightarrow 1^{-}}\alpha(\lambda,q).
\end{equation}
In Appendix~\ref{limitq} it is shown that the limit in~\eqref{eq.45n} can be analytically computed, resulting in the
$\lambda$-dependent storage capacity
\begin{equation}
\label{result0}
\alpha_c(\lambda)=\Bigg(\sum_{k=0}^{+\infty}\int_0^{2\pi/\lambda}\frac{{\rm d}\omega}{\sqrt{2\pi}}\,\omega^2\,\exp\Bigg(-\frac{1}{2}\Big(\omega+\frac{2\pi}{\lambda}k\Big)^2\Bigg)\Bigg)^{-1}\ .
\end{equation}
Notice that the series is uniformly convergent in
{ $\lambda>0$ (see Appendix \ref{series})}
then,  when $\lambda\to0^+$, only the contribution $k=0$ survives, so that
\begin{equation}
\label{result1}
\alpha_c(0)=\lim_{\lambda\to0^+}{\alpha_c(\lambda)}=\bigg(\int_0^{+\infty}\frac{{\rm d}\omega}{\sqrt{2\pi}}\,\omega^2\,{\rm e}^{-\frac{\omega^2}{2}}\bigg)^{-1}=2\ ,
\end{equation}
and one retrieves the standard storage capacity.
Furthermore, the function $\alpha_c(\lambda)$ is finite and infinitely differentiable at $\lambda=0$; however, all derivatives vanish at that point so that the storage capacity is not analytic there.
On the other hand, $\lim_{\lambda\to+\infty}\alpha_c(\lambda)=+\infty$ (see also Appendix~\ref{app.5}).

\begin{figure}[H]
    \begin{minipage}{0.47\textwidth}
        \centering
        \includegraphics[width=\textwidth]{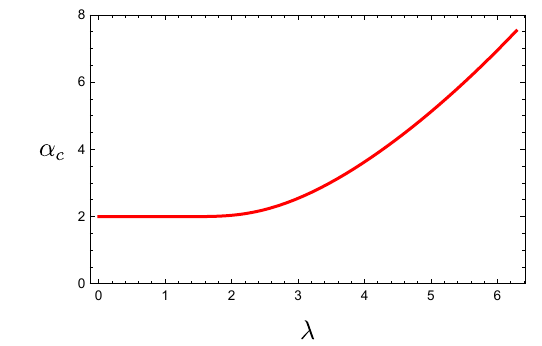}
    \end{minipage}%
    \hfill
    \begin{minipage}{0.5\textwidth}
        \centering
        \includegraphics[width=\textwidth]{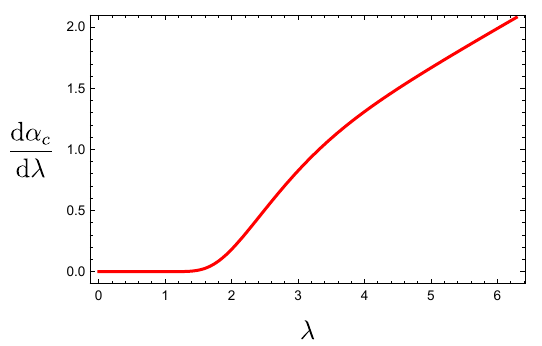}
    \end{minipage}
    \caption{Left: Plot of the storage capacity $\alpha$ in~\eqref{result0} as a function of $\lambda$. Right: First derivative of the storage capacity $\alpha$ in~\eqref{result0}with respect to $\lambda$.}
    \label{fig.1}
\end{figure}

In all cases, $\alpha_c(\lambda)\geq2$.
Indeed, changing the integration variable to  $\displaystyle t=\frac{\omega}{\sqrt{2}}+\frac{\sqrt{2}\pi}{\lambda}k$, one obtains
\begin{align}\nonumber
\sum_{k=0}^{+\infty}\int_0^{2\pi/\lambda} & \frac{{\rm d}\omega}{\sqrt{2\pi}}\,\omega^2\, \exp\Bigg(-\frac{1}{2}\Big(\omega+\frac{2\pi}{\lambda}k\Big)^2\Bigg)=
2\sum_{k=0}^{+\infty}\int_{\sqrt{2}\pi k/\lambda}^{\sqrt{2}\pi(k+1)/\lambda}\frac{{\rm d}t}{\sqrt{\pi}}\,\left(t-\frac{\sqrt{2}\pi k}{\lambda}\right)^2\,{\rm e}^{-t^2}\\ \nonumber
& =2 \int_0^{+\infty}\frac{{\rm d}t}{\sqrt{\pi}}\,t^2\,{\rm e}^{-t^2} + 2 \sum_{k=0}^{+\infty}\int_{\sqrt{2}\pi k/\lambda}^{\sqrt{2}\pi(k+1)/\lambda}\frac{{\rm d}t}{\sqrt{\pi}}\,\left(\frac{2\pi^2\,k^2}{\lambda^2}-\frac{2\sqrt{2}\pi t}{\lambda}\right)\,{\rm e}^{-t^2}\\
& =\frac{1}{2} + 2 \sum_{k=0}^{+\infty}\int_{\sqrt{2}\pi k/\lambda}^{\sqrt{2}\pi(k+1)/\lambda}\frac{{\rm d}t}{\sqrt{\pi}}\,\frac{2\pi k}{\lambda}\left(\frac{\pi k}{\lambda}-\sqrt{2}\,t\right)\,{\rm e}^{-t^2}\leq\frac{1}{2},
\label{result2}
\end{align}
where in the last line we used the fact that for $t\geq \sqrt{2}\pi k/\lambda$ one has
\begin{equation}
\frac{\pi k}{\lambda}-\sqrt{2}\,t\leq -\frac{\pi k}{\lambda}\leq 0\ .
\end{equation}
Inserting the bound~\eqref{result2} into~\eqref{result0} one concludes that
\begin{equation}
    \alpha_c(\lambda)\geq 2 \quad \text{ for all }\lambda>0.
\end{equation}
The left panel of
Figure~\ref{fig.1} illustrates the behavior of $\alpha$ in~\eqref{result0} as a function of $\lambda$. The value of $\alpha$ remains nearly constant at value $2$ until $\lambda$ approaches $2$, beyond which it increases monotonically and without bound. The right panel, instead,  presents the derivative of $\alpha$ with respect to $\lambda$, plotted against $\lambda$. The derivative remains close to zero up to $\lambda \approx 1.6$, after which it begins to increase gradually and monotonically. An evidence of the fact that $\alpha$ increases nonlinearly with $\lambda$.

\section{Conclusions and outlook}
Summarizing, building on the discrete model presented in~\cite{torr2019} for implementing a quantum perceptron, we modified the unitary gate~\eqref{eq.41s} to explore how variations in the oscillation period affect the system's behavior. Then, using the replica method, we analytically computed the storage capacity as a function of the oscillation frequency $\lambda$. In the limit of vanishing frequency, the classical value of $2$ is recovered. However, as the frequency increases, we observe an indefinite enhancement in the quantum storage capacity beyond the classical threshold. This behavior is as expected taking into account that the periodicity of the activation function provides a “finer partitioning” of the input-state space, thereby leading to an increased perceptron storage capacity. However, as the frequency rises, this advantage may turn into a liability, since overfitting tends to take over.
This kind of drawback can be ascertained via a thorough study of the generalization error in a teacher-student setting, where explicit overfitting effects are expected to emerge and to be accessible by means of Gardner's approach. We plan to address this intriguing problem in a future work together with the robustness of our results in the presence of noise in the unitary operations. This kind of noise can be mimicked by means of either normal distributed random weights or inputs in a statistical learning scenario~\cite{enge2000}.
Furthermore, it is also important to explore how these results are affected by a replica symmetry breaking, for example within the one-step replica symmetric ansatz.

As a general remark, the observed improvement of the storage capacity beyond the classical threshold $\alpha=2$ arises solely from the specific form of the activation function. This indicates that, in principle, comparable enhancements could be achieved within a fully classical framework. Supporting evidence appears in~\cite{sopena}, where the authors argue that non‑monotonic activation functions can substantially improve neural‑network performance. In particular, they show that employing a sine activation function in a feedforward architecture enhances both generalization and learning speed relative to sigmoid activations, thereby providing a classical benchmark for our findings.
Actually, in the quantum perceptron model considered here, the nonlinearity of the activation function does not stem from an intrinsic quantum mechanism. Instead, it emerges from taking expectation values of a measured observable—a procedure that is effectively classical. In this sense, the nonlinearity is “measure‑induced” through averaging and could, in principle, be reproduced within a classical setting.
By contrast, a genuinely quantum source of nonlinearity could arise, for example, from a selective measurement process in which the perceptron’s output is determined directly by a single measurement outcome rather than by an averaged expectation value. Such a mechanism would introduce a form of nonlinearity that cannot be reproduced classically. Another potential route to quantum advantage may emerge at the level of a full neural network, where interference between different perceptrons becomes relevant and the resulting performance gains stem from intrinsically quantum interference effects.
\appendix

\section{Computation of the quantum storage capacity~\texorpdfstring{\eqref{eq.45n}}{}}
\label{app.1}

We now compute the expression of the storage capacity $\alpha_c(\lambda)$ of Subsection~\ref{subsec.3.1}. We consider a dataset of the form $\{\x^\mu,\xi^\mu\}_{\mu=1}^p$, where $\x^\mu\in\{-1,1\}^N$, while $\xi^\mu=\pm 1$ is a binary label. Using the parity of $\sin(z)$, the Gardner volume~\eqref{eq.44s} can be rewritten as
\begin{align}
 V_N^\lambda(\{\x^\mu,\xi^\mu\}_{\mu=1}^p)
&=\int\de\mu(\bm{w})\prod_{\mu=1}^p\Theta\left(\sin\left(\lambda\frac{\w\cdot\bm{r}^\mu}{\sqrt{N}}\right)\right), \qquad \bm{r}^\mu=\xi^\mu\x^\mu.
\end{align}
The quantity of interest is the expectation value of $\ln V$ w.r.t.\ the distribution of patterns and the labels. Note that the distribution~\eqref{eq:distPatternsLabels} on $\{\x^\mu,\xi^\mu\}_{\mu=1}^p$ induces the distribution
\begin{equation}\label{eq:ProbR}
    P(r^{\mu}_j=1)=P(r^{\mu}_j=-1)=\frac{1}{2}.
\end{equation}
 In the following we will denote with $\langle\cdot \rangle$ expectations with respect to the distribution~\eqref{eq:ProbR}. The computation of $\langle\ln V_N^\lambda\rangle$ {is performed} by the replica trick~\eqref{eq:replicaTrick}, which leads us to consider the following expectation value
\begin{equation}\label{eq.55}
\langle (V_N^\lambda)^n\rangle=\int \prod_{ \gamma =1}^n \de\mu(\w_\gamma )\prod_{\mu=1}^p\bigg\langle\prod_{ \gamma =1}^n\Theta\left(\sin\left(\lambda\frac{\w_\gamma \cdot\bm{r}^\mu}{\sqrt{N}}\right)\right)\bigg\rangle.
\end{equation}
Using the spin glass order parameters
\begin{equation}
    \label{eq:overlap}
q_{ \gamma \delta}\coloneq\frac{1}{N}\w_\gamma \cdot\w_\delta\ ,
\end{equation}
the integral~\eqref{eq.55} can be rewritten as
\begin{equation}\label{eq.58}
\langle (V_N^\lambda)^n\rangle=\frac{1}{C_{N}^n}\int\de Q\int\de\bm{W}\, \delta\left(\bm{W}^T\bm{W}-NQ\right)\prod_{\mu=1}^p\bigg\langle\prod_{ \gamma =1}^n\Theta\left(\sin\left(\lambda\frac{\w_\gamma \cdot\bm{r}^\mu}{\sqrt{N}}\right)\right)\bigg\rangle,
\end{equation}
where, we introduced the $N\times n$ matrix $\bm{W}=(\bm{w}_1,\dots,\bm{w}_n)$ whose columns are $\w_\gamma $, the $n\times n$ matrix $Q$ whose elements are $q_{ \gamma \delta}$ and we introduced the notations
\begin{equation}
    \de \bm{W}\coloneq\prod_{ \gamma =1}^n\de\w_\gamma , \qquad \de Q\coloneq\prod_{\gamma <\delta}\,\de q_{ \gamma \delta},
\end{equation}
\begin{equation}
    \delta(\bm{W}^T\bm{W}-NQ)\coloneq\prod_{ \gamma\leq\delta}\delta(\w_\gamma \cdot\w_\delta-NQ).
\end{equation}
Note that the integration in $Q$ is done only over the upper off-diagonal terms, since the matrix is symmetric by definition and the diagonal terms are fixed to $q_{\gamma\gamma}=1$ by the normalization condition $\w_\gamma \cdot \w_\gamma  =N$.
A standard computation (see section~\ref{sec:energyCalc} for the details) shows that  to the leading order in the thermodynamic limit $N\rightarrow\infty$:
\begin{equation}
    \label{eq:ExpEnergy}
    \prod_{\mu=1}^p\bigg\langle\prod_{ \gamma =1}^n\Theta\left(\sin\left(\lambda\frac{\w_\gamma \cdot\bm{r}^\mu}{\sqrt{N}}\right)\right)\bigg\rangle\delta\left(\bm{W}^T\bm{W}-NQ\right)\simeq {\rm e}^{NE_\lambda(Q)}\delta\left(\bm{W}^T\bm{W}-NQ\right),
\end{equation}
with
\begin{equation}\label{eq:Energy}
    E_\lambda(Q)=\alpha\ln\left(\frac{1}{(2\pi)^n}\int_{\Sigma^{n}}\de\bm{z} \int_{\R^n}\de \bm{y}\, \rme ^{\ii \bm{z}\cdot \bm{y}-\frac{\lambda^2}{2}\bm{y}\cdot Q\bm{y}}\right),
\end{equation}
where $\bm{z}\coloneq(z_1,\dots,z_n)$, $\bm{y}\coloneq(y_1,\dots,y_n)$ and
\begin{equation}
    \Sigma\coloneq\{z\in\mathbb{R}\,|\, 2k\pi \leq z \leq (2k+1)\pi, k\in\Z\}.
\end{equation}
In other words, in the thermodynamic limit only the overlap matrix $Q$
is relevant, while the remaining degrees of freedom in  $\bm{W}$ are redundant. This redundancy can be integrated out, considering
\begin{equation}
    {\rm e}^{NS(Q)}= \int \de\bm{W}\delta(\bm{W}^T\bm{W}-QN),
    \label{eq:expS}
\end{equation}
which represents the volume in the $\bm{W}$ space consistent with the constraints~\eqref{eq:overlap}.
The details of this change of variables, including the evaluation of the Jacobian determinant involved, are worked out explicitly in~\cite{Fyodorov}. The result, up to irrelevant constants, is given by
\begin{equation}
\rme^{NS(Q)}\propto\det(Q)^{\frac{N-n-1}{2}}.
\end{equation}
Therefore, to the leading order in $N\rightarrow\infty$:
\begin{equation}\label{eq:S(Q)}
    S(Q)\simeq \frac{1}{2}\ln\det(Q).
\end{equation}
Using~\eqref{eq:ExpEnergy} and~\eqref{eq:expS}, equation~\eqref{eq.58} can be recast as
\begin{equation}\label{eq:VnCompact}
\langle (V_N^\lambda)^n\rangle=\int\de Q \, {\rm e}^{ N[E_\lambda(Q)+S(Q)]}.
\end{equation}
Note that $E_\lambda(Q)$ is the only term containing the information on the specific problem we are considering (through the integration domain $\Sigma$), while the term $S(Q)$ is purely geometric in nature. Therefore they are usually called the energetic and entropic contribution, respectively.
Expression~\eqref{eq:VnCompact} is well suited for a computation of the replicated volume in the thermodynamic limit $N\rightarrow \infty$ through a saddle-point approximation:
\begin{equation}
    \frac{\ln\langle (V_N^\lambda)^n\rangle}{N}=\underset{Q}{\mathrm{extr}}[E_\lambda(Q)+S(Q)].
\end{equation}
 \subsection{Replica symmetric ansatz}
In the replica symmetric ansatz, namely
$q_{ \gamma \delta}=q$ for each $\gamma\neq\delta$,~\eqref{eq:Energy} takes the form
\begin{equation}
E_\lambda(Q)=\alpha\ln\left(\frac{1}{(2\pi)^n}\int_{\Sigma^{n}}\de\bm{z} \int_{\mathbb{R}^n}\de \bm{y}\, \rme ^{\ii \bm{z}\cdot \bm{y}-\frac{\lambda^2}{2}(1-q)\|\bm{y}\|^2}\rme^{-\frac{\lambda^2}{2}q\left(\sum_{ \gamma =1}^ny_\gamma \right)^2}\right).
\end{equation}
The last exponential can be dealt with the introduction of an additional auxiliary Gaussian variable through the Hubbard-Stratonovich transformation
\begin{equation}
    \rme^{-\frac{a}{2}\xi^2}=\int_\mathbb{R} D\omega\,\rme^{-\ii \sqrt{a}\xi \omega}, \qquad \text{where}\quad D\omega\coloneq\de\omega\frac{\rme^{-\frac{\omega^2}{2}}}{\sqrt{2\pi}},
\end{equation}
which yields factorization over the replicas:
\begin{align}
    \notag E_\lambda(Q)&=\alpha\ln\left(\int_\mathbb{R} D\omega\left(\frac{1}{2\pi}\int_{\Sigma}\de z \int_{\R}\de y\, \rme^{\ii ( z-\sqrt{q}\lambda) y-\frac{\lambda^2}{2}(1-q)y^2}\right)^n\right)\\
    &=\alpha\ln\left(\int_\mathbb{R} D\omega\left(\frac{1}{[2\pi \lambda^2(1-q)]^{1/2}}\int_{\Sigma}\de z\, \rme^{-\frac{(z-\sqrt{q}\lambda)^2}{2\lambda^2(1-q)}}\right)^n\right).
\end{align}
After a change of variables, we get to the expression
\begin{equation}\label{eq:EnergyRS}
    E_\lambda(Q)=\alpha \ln\left(\int_\mathbb{R} D\omega\left(\int_{\Sigma_\lambda^q}Dz\right)^n\right),
\end{equation}
where
\begin{equation}\label{eq:lambdaQdomain}
\Sigma^\lambda_{q}\coloneqq\bigcup_{k\in\Z}\Big[\frac{-\lambda\sqrt{q}\omega+2k\pi}{\lambda\sqrt{1-q}},\frac{-\lambda\sqrt{q}\omega+(2k+1)\pi}{\lambda\sqrt{1-q}}\Big]\ .
\end{equation}
The leading order of~\eqref{eq:EnergyRS} in the $n\rightarrow0$ limit is given by
\begin{equation}
    E_\lambda(Q)\simeq n\alpha \int_\mathbb{R} D\omega\ln\left(\int_{\Sigma_\lambda^q}Dz\right)=n\alpha \int_\mathbb{R} D\omega\ln\Psi(\lambda,q,\omega)
\end{equation}
where we set
\begin{equation}
\label{eq:Psi}
    \Psi(\lambda,q,\omega)\coloneqq \sum_{k\in\Z}\int_{\epsilon_1(\lambda,q,\omega,k)}^{\epsilon_2(\lambda,q,\omega,k)}\frac{\de z}{\sqrt{2\pi}}\,{\rm e}^{-\frac{z^2}{2}},
\end{equation}
with
\begin{equation}
\epsilon_1(\lambda,q,\omega,k)\coloneqq \frac{-\lambda\sqrt{q}\omega+2k\pi}{\lambda\sqrt{1-q}},\qquad\epsilon_2(\lambda,q,\omega,k)\coloneqq \frac{-\lambda\sqrt{q}\omega+(2k+1)\pi}{\lambda\sqrt{1-q}}.
\end{equation}
To evaluate the entropic term within the replica symmetric ansatz, it is useful to note that the eigenvalue spectrum of the overlap matrix  $Q$ can be obtained explicitly: $n-1$ eigenvalues are equal to $(1 - q)$, and a single eigenvalue is $(1 + q(n - 1))$. The determinant of $Q$ thus reads
\begin{equation}
  \det Q = (1 + q(n - 1))(1 - q)^{n - 1}.
\end{equation}
Consequently, to the leading order in $n\rightarrow 0$:
\begin{align}
    S(Q)&=\frac{1}{2}\ln\det Q\\
    &=\frac{1}{2}\left[\ln\left(1+q(n-1)\right) +(n-1)\ln(1-q)\right]\\
    &=\frac{n}{2}\left[\frac{q}{1-q}+\ln(1-q)\right]+O(n^2).
\end{align}
Therefore, in the replica symmetric ansatz one has
\begin{equation}\label{eq:FreeEnG}
\frac{\ln\langle (V_N^\lambda)^n\rangle}{nN}=\underset{0\leq q\leq 1}{\mathrm{extr}}\mathcal{G}(\lambda,\alpha,q),
\end{equation}
where
\begin{equation}
\mathcal{G}(\lambda,\alpha,q)\coloneq\alpha \int_\mathbb{R} D\omega\ln\Psi(\lambda,q,\omega)+\frac{1}{2}\left[\frac{q}{1-q}+\ln(1-q)\right].
\end{equation}
The saddle-point equation $\partial_q\mathcal{G}(\lambda,\alpha,q)=0$  then reads
\begin{equation}
    \alpha\int D\omega \frac{\partial}{\partial q}\ln\Psi(\lambda,q,\omega)+\frac{q}{2(1-q)^2}=0,
\end{equation}
which finally yields
\begin{equation}
\alpha(\lambda,q)
=-\frac{q}{2(1-q)^2}\bigg(\int\frac{\de \omega}{\sqrt{2\pi}}\,{\rm e}^{-\frac{\omega^2}{2}}\frac{\Phi(\lambda,q,\omega)}{\Psi(\lambda,q,\omega)} \bigg)^{-1},
\end{equation}
where we wrote explicitly the gaussian measure $D\omega$ and we set
\begin{equation}
\label{eq:Phi}
\Phi(\lambda,q,\omega)\coloneq \partial_q\Psi(\lambda,q,\omega).
\end{equation}

\subsection{Derivation of \texorpdfstring{$E_\lambda(Q)$}{}}\label{sec:energyCalc}
In this section we present the detailed derivation of equations~\eqref{eq:ExpEnergy}-\eqref{eq:Energy}.

First, note that for $\eta\in\mathbb{R}$
\begin{equation}
    \Theta(\sin(\eta))=
    \begin{cases}
        1 \quad &\mathrm{if } \ \eta\in\Sigma\\
        0 &\mathrm{otherwise}
    \end{cases},
    \qquad \Sigma=\{\eta\in\mathbb{R}\,|\, 2k\pi\leq\eta\leq (2k+1)\pi,\ k\in\Z\}
\end{equation}
admits the integral representation
\begin{equation}
    \Theta(\sin(\eta))=\int_{\Sigma}\de z \, \delta(z-\eta)=\frac{1}{2\pi}\int_{\Sigma}\de z \, \int_\mathbb{R} \de y\, \rme^{\ii y(z-\eta)}.
\end{equation}
Therefore:
\begin{align}\notag
&\prod_{\mu=1}^p\bigg\langle\prod_{ \gamma =1}^n\Theta\left(\sin\left(\lambda\frac{\w_\gamma \cdot\bm{r}^\mu}{\sqrt{N}}\right)\right)\bigg\rangle\\
&\qquad=\int_{\Sigma^{np}}\de Z \int_{\mathbb{R}^{np}}\de Y\frac{1}{(2\pi)^{np}}  \, \rme^{\ii Z\cdot Y}\prod_{\mu=1}^p\bigg\langle \exp\left(-\ii\frac{1}{\sqrt{N}} \sum_{ \gamma =1}^ny_{\gamma}^\mu \w_\gamma \cdot\bm{r}^\mu\right)\bigg \rangle,
\label{eq:Average0}
\end{align}
where we introduced the short-hand notation $Z\coloneq (z_{ z }^{\mu})_{ z ,\mu}$, $Y\coloneq(y_\gamma ^\mu)_{ z ,\mu}$. Using the fact that the components of $\bm{r}^\mu$ are independent and distributed according to~\eqref{eq:ProbR}, we get
\begin{align}\notag
\bigg\langle \exp\bigg(-\ii \frac{1}{\sqrt{N}}\sum_{ \gamma =1}^ny_{\gamma}^\mu \w_\gamma \cdot\bm{r}^\mu\bigg)\bigg \rangle&=\prod_{j=1}^N\bigg\langle \exp\bigg(-\ii \frac{1}{\sqrt{N}}\sum_{ \gamma =1}^ny_{\gamma}^\mu w_{\gamma j}r^\mu_j\bigg)\bigg \rangle\\
&=\prod_{j=1}^N \cos\bigg(\frac{1}{\sqrt{N}} \sum_{ \gamma =1}^n y_{\gamma}^\mu w_{\gamma j}\bigg).
\label{eq:Average}
\end{align}
Now note that to the leading order in $N\rightarrow\infty$:
\begin{align}\notag
\prod_{j=1}^N \cos\bigg(\frac{1}{\sqrt{N}} \sum_{ \gamma =1}^ny_{\gamma}^\mu w_{\gamma j}\bigg)&=\exp\left(\sum_{j=1}^N\ln\left[\cos\bigg(\frac{1}{\sqrt{N}} \sum_{ \gamma =1}^ny_{\gamma}^\mu w_{\gamma j}\bigg)\right]\right)\\\notag
&\simeq \exp\left(-\frac{1}{2N}\sum_{j=1}^{N}\sum_{\gamma, \delta =1}^{n}y_\gamma ^\mu y_\delta^\mu w_{\gamma j} w_{\delta j}\right)\\
&=\exp\left(-\frac{1}{2}\sum_{\gamma, \delta =1}^{n}y_\gamma ^\mu y_\delta^\mu q_{ \gamma \delta}\right),
\label{eq:AverageLeading}
\end{align}
where we used the expansion $\ln(\cos(\varepsilon))\simeq -\varepsilon^2/2$ as $\varepsilon\rightarrow 0$ and the definition~\eqref{eq:overlap}. Insertion of~\eqref{eq:AverageLeading} into~\eqref{eq:Average} allows to rewrite~\eqref{eq:Average0} as
\begin{align}\notag
&\prod_{\mu=1}^p\bigg\langle\prod_{ \gamma =1}^n\Theta\left(\sin\left(\lambda\frac{\w_\gamma \cdot\bm{r}^\mu}{\sqrt{N}}\right)\right)\bigg\rangle \\ \notag
&\qquad=\int_{\Sigma^{np}}\de Z \int_{\mathbb{R}^{np}}\de Y\frac{1}{(2\pi)^{np}}  \, \prod_{\mu=1}^p\exp\left(\ii\sum_{ \gamma =1}^n z^{\mu}_\gamma    y^\mu_\gamma   -\frac{1}{2}\sum_{\gamma, \delta =1}^{n}y_\gamma ^\mu y_\delta^\mu q_{ \gamma \delta}\right)\\ \notag
&\qquad =\left(\frac{1}{(2\pi)^n}\int_{\Sigma^{n}}\de\bm{z} \int_{\mathbb{R}^n}\de \bm{y}\, \rme ^{\ii \bm{z}\cdot \bm{y}-\frac{\lambda^2}{2}\bm{y}\cdot Q\bm{y}}\right)^p\\
    &\qquad=\rme^{NE_\lambda(Q)},
\end{align}
with $E_\lambda(Q)$ given in~\eqref{eq:Energy}.
\section{Computation of  \texorpdfstring{$\lim_{q\to1^-}\alpha(\lambda,q)$}{}}\label{limitq}

In order to handle the function $\Psi(\lambda,q,\omega)$ in~\eqref{eq.46s}, we write
\begin{equation}
{\rm e}^{-\frac{ z ^2}{2}}=\int_{-\infty}^{+\infty}\de y\, \frac{{\rm e}^{\ii  y  z }}{\sqrt{2\pi}}\,{\rm e}^{-y^2/2};
\end{equation}
then, integrating
\begin{align}\nonumber
\int_{\epsilon_1(\lambda,q,\omega,k)}^{\epsilon_2(\lambda,q,\omega,k)}\frac{\de z }{\sqrt{2\pi}}\,\frac{{\rm e}^{\ii  y  z }}{\sqrt{2\pi}} &=
\frac{1}{2\pi \ii y}\left({\rm e}^{\ii  y \frac{(2k+1)\pi-\lambda\sqrt{q}\omega}{\lambda\sqrt{1-q}}}-{\rm e}^{\ii  y \frac{2k\pi-\lambda\sqrt{q}\omega}{\lambda\sqrt{1-q}}}\right)
\\
&={\rm e}^{\ii  y \frac{2 k \pi}{\lambda\sqrt{1-q}}}\,e^{-\ii y\frac{\sqrt{q}\omega}{\sqrt{1-q}}}\,
\frac{{\rm e}^{\ii \frac{ y \pi}{\lambda\sqrt{1-q}}}\,-\,1}{2 \pi \ii y}\ .
\end{align}
The partial sums over $-n\leq k\leq n$ provide a sequence of \textit{Dirichlet kernels}
\begin{equation}
\sum_{k=-n}^n{\rm e}^{\ii  y \frac{2 k \pi}{\lambda\sqrt{1-q}}}=\frac{\sin\left(\frac{2\pi y}{\lambda\sqrt{1-q}}(n+\frac{1}{2})\right)}{\sin\left(\frac{\pi y}{\lambda\sqrt{1-q}}\right)}\eqqcolon {\rm D}_n\Big(\frac{2\pi y}{\lambda\sqrt{1-q}}\Big).
\end{equation}
Sending $n\to+\infty$, one gets the so-called \textit{Dirac comb} distribution:
\begin{equation}
\lim_{n\to+\infty}{\rm D}_n\Big(\frac{2\pi y}{\lambda\sqrt{1-q}}\Big)=2\pi\sum_{k=-\infty}^{+\infty}\delta\Big(\frac{2\pi y}{\lambda\sqrt{1-q}}-2k\pi\Big)=\lambda\sqrt{1-q}\sum_{k=-\infty}^{+\infty}\delta\Big(y-\lambda k\sqrt{1-q}\Big),
\end{equation}
which, inserted into~\eqref{eq.46s} finally yields
\begin{align} \nonumber
\Psi(\lambda,q,\omega)&=\frac{1}{2}+\sum_{k=1}^{+\infty}\frac{{\rm e}^{-(1-q)\lambda^2 k^2/2}}{\pi k}\big(1-(-1)^k\big)\,\sin\big(k\lambda\sqrt{q}\omega\big) \\ \label{Psi}
&=\frac{1}{2}+\frac{2}{\pi}\sum_{m=0}^{+\infty}\frac{{\rm e}^{-(1-q)\lambda^2(2m+1)^2/2}}{2m+1}\,\sin\big((2m+1)\lambda\sqrt{q}\omega\big).
\end{align}

To compute the critical storage capacity as a function of $\lambda$, we need to compute the limit
\begin{equation}
\label{eq0}
\alpha(\lambda)=\lim_{q\to1^-}\alpha(\lambda,q)=-\frac{1}{2}\Big(\lim_{q\to1^-}(1-q)^2\,I(\lambda,q)\Big)^{-1},
\end{equation}
where we set
\begin{equation}
\label{eq0.1}
I(\lambda,q)\coloneq\int_{-\infty}^{+\infty}\frac{\de \omega}{\sqrt{2\pi}}\,{\rm e}^{-\frac{\omega^2}{2}}
\frac{\partial_q\Psi(\lambda, q,\omega)}{\Psi(\lambda, q,\omega)}.
\end{equation}
Let us consider
\begin{equation}\label{J-integral}
J(\lambda,q)\coloneq(1-q)^2\,I(\lambda,q)=(1-q)^2\,\int_{-\infty}^{+\infty}\frac{\de \omega}{\sqrt{2\pi}}\,{\rm e}^{-\frac{\omega^2}{2}}
\frac{\partial_q\Psi(\lambda, q,\omega)}{\Psi(\lambda, q,\omega)}.
\end{equation}
By differentiating~\eqref{Psi} term by term, we get
\begin{align}\nonumber
\Phi(\lambda,q,\omega)\coloneq\partial_q\Psi(\lambda, q,\omega)=&\frac{1}{\pi}\sum_{m=0}^{+\infty}{\rm e}^{-(1-q)\lambda^2(2m+1)^2/2}\,\Big(\lambda^2(2m+1)\sin\big((2m+1)\lambda\sqrt{q}\omega\big) \\ \label{eq2a}
& \hspace{4.5cm} +\frac{\lambda\omega}{\sqrt{q}}\cos\big((2m+1)\lambda\sqrt{q}\omega\big)\Big).
\end{align}
In order to deal with $J(\lambda,q)$ in \eqref{J-integral},
notice that, for all integers $n\in\mathbb{Z}$, one has
\begin{align}\label{Per1}
\Psi\big(\lambda, q,\omega\pm nT\big)&=\Psi(\lambda, q,\omega),\\\label{Per2}
\Phi\big(\lambda, q,\omega\pm nT\big)&=\Phi(\lambda, q,\omega)\pm\, n\,\mathcal{X}(\lambda,q,\omega),
\end{align}
where $T\coloneq\frac{2\pi}{\lambda\sqrt{q}}$, and
\begin{align}\label{Per3}
\mathcal{X}(\lambda,q,\omega)&\coloneq\frac{2}{q}\sum_{m=0}^{+\infty}{\rm e}^{-(1-q)\lambda^2(2m+1)^2/2}\,\cos\big((2m+1)\lambda\sqrt{q}\omega\big), \\
\mathcal{X}\big(\lambda,q,\omega\pm nT\big)&=\mathcal{X}(\lambda,q,\omega).
\end{align}
One then rewrites
\begin{align}\nonumber
J(\lambda,q) =& (1-q)^2\, \sum_{n=0}^{+\infty} \left(\int_{nT}^{(n+1)T}+\int_{-(n+1)T}^{-nT}\right)\frac{\de \omega}{\sqrt{2\pi}}\,{\rm e}^{-\frac{\omega^2}{2}}
\,\frac{\Phi(\lambda, q,\omega)}{\Psi(\lambda, q,\omega)}\\
\nonumber
=& (1-q)^2\, \sum_{n=0}^{+\infty} \Bigg(\int_0^{T}\frac{\de \omega}{\sqrt{2\pi}}\,{\rm e}^{-(\omega+nT)^2/2}\,\frac{\Phi(\lambda, q,\omega)+n\mathcal{X}(\lambda,q,\omega)}{\Psi(\lambda, q,\omega)}\\ \label{result3a}
& \hspace{2cm} + \int_{-T}^0\frac{\de \omega}{\sqrt{2\pi}}\,{\rm e}^{-(\omega-nT)^2/2}\,\frac{\Phi(\lambda, q,\omega)-n\mathcal{X}(\lambda,q,\omega)}{\Psi(\lambda, q,\omega)}\Bigg)\ .
\end{align}
Changing $\omega$ into $-\omega$ in the second integral yields
\begin{align}\nonumber
J(\lambda,q) =& (1-q)^2\, \sum_{n=0}^{+\infty} \int_0^{T} \frac{\de \omega}{\sqrt{2\pi}}\,{\rm e}^{-(\omega+nT)^2/2} \\ \label{result3c}
& \hspace{0.5cm}\times\left[
\frac{\Phi(\lambda, q,\omega)}{\Psi(\lambda, q,\omega)}+\frac{\Phi(\lambda, q,-\omega)}{\Psi(\lambda, q,-\omega)}+n\mathcal{X}(\lambda,q,\omega)\bigg(
\frac{1}{\Psi(\lambda, q,\omega)}-\frac{1}{\Psi(\lambda, q,-\omega)}\bigg)\right].
\end{align}

Now, by changing the integration variable from $\omega$ to $\nu=\omega/\sqrt{1-q}$, we get
\begin{align}\nonumber
J(\lambda,q)=&(1-q)^{5/2}\,\sum_{n=0}^{+\infty}\int_0^{2\pi/(\lambda\sqrt{q(1-q)})}\frac{\de \nu}{\sqrt{2\pi}}\,{\rm e}^{-\frac{1}{2}\big(\nu\sqrt{1-q}+2\pi n/(\lambda\sqrt{q})\big)^2} \\ \nonumber
&  \times\left[
\frac{\Phi(\lambda, q,\nu\sqrt{1-q})}{\Psi(\lambda, q,\nu\sqrt{1-q})}+\frac{\Phi(\lambda, q,-\nu\sqrt{1-q})}{\Psi(\lambda, q,-\nu\sqrt{1-q})}
\right.\\ \label{eq2c}
&\hspace{2cm}  +\left.n\mathcal{X}(\lambda,q,\nu\sqrt{1-q})\bigg(
\frac{1}{\Psi(\lambda, q,\nu\sqrt{1-q})}-\frac{1}{\Psi(\lambda, q,-\nu\sqrt{1-q})}\bigg)\right].
\end{align}

Setting $x_m\coloneq(2m+1)\sqrt{1-q}$, so that $\Delta x_m=x_{m+1}-x_m=2\sqrt{1-q}$, we can recast
\begin{align}\label{eq3a}
\Psi(\lambda, q,\nu\sqrt{1-q})=&\frac{1}{2}+\frac{1}{\pi}\sum_{m=0}^{+\infty}\Delta x_m\,\frac{{\rm e}^{-\lambda^2x_m^2/2}}{x_m}\,\sin\big(x_m\lambda\nu\sqrt{q}\big),\\
\nonumber
\Phi(\lambda, q,\nu\sqrt{1-q})=&\frac{\lambda^2}{2\pi(1-q)}\sum_{m=0}^{+\infty}\Delta x_m\, {\rm e}^{-\lambda^2x_m^2/2}\, x_m\sin\big(x_m\lambda\nu\sqrt{q}\big)\\ \label{eq3b}
&{\color{red}}+\frac{\lambda\nu}{2\pi\sqrt{q}}\sum_{m=0}^{+\infty}\Delta x_m\,{\rm e}^{-\lambda^2x_m^2/2}\, \cos\big(x_m\lambda\nu\sqrt{q}\big),\\
\label{eq3c}
\mathcal{X}(\lambda,q,\nu\sqrt{1-q})=&\frac{1}{q\sqrt{1-q}}\sum_{m=0}^{+\infty}\Delta x_m\,{\rm e}^{-\lambda^2x_m^2/2}\,
\cos\big(x_m\lambda\nu\sqrt{q}\big).
\end{align}
Notice that the discrete points $x_m$ are in the middle of the intervals
\begin{equation}
\Big[2m\sqrt{1-q}\,,\,2(m+1)\sqrt{1-q}\Big].
\end{equation}
It then follows that the series in~\eqref{eq3a}-\eqref{eq3c} are Riemann sums such that, when $q\to1^-$,
\begin{align}\label{eq6a}
\Psi\big(\lambda, q,\nu\sqrt{1-q}\big) &\simeq
\frac{1}{2}{+}\frac{1}{\pi}\int_0^{+\infty} \de y\,{\rm e}^{-y^2/2}\frac{\sin(y\nu)}{y}=\frac{1}{2}\big(1+\mathrm{erf}(\nu/\sqrt{2})\big),\\ \nonumber
\Phi\big(\lambda,q,\nu\sqrt{1-q}\big) &\simeq \frac{1}{2\pi(1-q)}\int_0^{+\infty} \de y\,{\rm e}^{-y^2/2}\,y\,\sin(y\nu){\color{red}}+\frac{\nu}{2\pi}\int_0^{+\infty} \de y\,{\rm e}^{-y^2/2}\,\cos(y\nu) \\ \label{eq6b}
&\simeq\frac{\nu}{2\sqrt{2\pi}(1-q)}\,{\rm e}^{-\nu^2/2}\,, \\ \label{eq6c}
\mathcal{X}(\lambda,q,\nu\sqrt{1-q})&\simeq\frac{1}{\lambda\sqrt{1-q}}\int_0^{+\infty}\de y\,{\rm e}^{-y^2/2}\,\cos(y\nu)=\frac{1}{\lambda}\sqrt{\frac{\pi}{2(1-q)}}\,\,{\rm e}^{-\nu^2/2}\,,
\end{align}
where, in passing from the first line to the second one in equation~\eqref{eq6b}, we have only retained the dominant diverging contribution
when $q\to1^-$ and, in~\eqref{eq6a}, we have introduced the error function
\begin{equation}\label{erf}
\mathrm{erf}(z)=\frac{2}{\sqrt{\pi}}\int_0^z\de t\,{\rm e}^{-t^2}.
\end{equation}
Then, as $q\to1^{-}$, the dominant behavior of $J(\lambda,q)$ in~\eqref{eq2c} is determined by~\eqref{eq6a} and~\eqref{eq6b}, the latter diverging faster than~\eqref{eq6c}, as $\displaystyle \frac{1}{1-q}$ against $\displaystyle\frac{1}{\sqrt{1-q}}$. Therefore,
\begin{align}\nonumber
J(\lambda,q) \simeq& (1-q)^{3/2} \sum_{n=0}^{+\infty} \int_0^{2\pi/(\lambda\sqrt{1-q})}\frac{\de\nu}{2\pi}\,\nu\,\exp\Big(-\frac{\nu^2}{2}-\frac{1}{2}\Big(\nu\sqrt{1-q}+\frac{2\pi n}{\lambda}\Big)^2\Big)\\
\label{result4}
&\hspace{3cm}
\times\Big(\frac{1}{1+\mathrm{erf}(\nu/\sqrt{2})}-\frac{1}{1-\mathrm{erf}(\nu/\sqrt{2})}\Big).
\end{align}
Going back to the integration variable $\omega=\nu\sqrt{1-q}$, one gets
\begin{align}\nonumber
J(\lambda,q) \simeq & (1-q)^{1/2}\, {\sum_{n=0}^{+\infty}} \int_{0}^{2\pi/\lambda}\frac{\de \omega}{2\pi}\,\omega\,\exp\Big(-\frac{\omega^2}{2(1-q)}-\frac{1}{2}\Big(\omega+\frac{2\pi n}{\lambda}\Big)^2\Big) \\
\label{eq8}
&\hspace{3cm}
\times\Big(\frac{1}{1+\mathrm{erf}(\omega/\sqrt{2(1-q)})}-\frac{1}{1-\mathrm{erf}(\omega/\sqrt{2(1-q)})}\Big).
\end{align}
From the asymptotic behavior
\begin{equation}
\label{eq9}
\mathrm{erf}(z)\simeq 1-\frac{1}{\sqrt{\pi}z}\,{\rm e}^{-z^2}\qquad\text{when}\qquad z\to+\infty\ ,
\end{equation}
one obtains the expression~\eqref{result0}; indeed,
\begin{align}\nonumber
J(\lambda,q) \simeq & (1-q)^{1/2}\, {\sum_{n=0}^{+\infty}} \int_{0}^{2\pi/\lambda}\frac{\de \omega}{2\pi}\,\omega\,\exp\Big(-\frac{\omega^2}{2(1-q)}-\frac{1}{2}\Big(\omega+\frac{2\pi n}{\lambda}\Big)^2\Big)\\
\nonumber
&\hspace{3cm} \times \bigg(\frac{1}{2}-\sqrt{\frac{\pi}{2(1-q)}}\,\omega\,\exp\Big(\frac{\omega^2}{2(1-q)}\bigg)\\
\label{result5b}
\simeq& -\sum_{n=0}^{+\infty}\int_0^{2\pi/\lambda}\frac{\de \omega}{2\sqrt{2\pi}}\,\omega^2\,\exp\bigg(-\frac{1}{2}\Big(\omega+\frac{2\pi n}{\lambda}\Big)^2\bigg).
\end{align}

{
\section{Uniform convergence of the series in~\texorpdfstring{\eqref{result0}}{}}\label{series}
}

Setting $\frac{1}{\lambda}\coloneqq s$, consider the sequence
\begin{equation}
\left\{ f_k(s)\coloneqq\int_0^{2\pi s} \frac{{\rm d}\omega}{\sqrt{2\pi}}  \omega^2 e^{-\frac{1}{2}(\omega+2\pi s k)^2}
\right\}_k.
\end{equation}
{
Taking the derivative with respect to $s$ we obtain
\begin{equation}
\frac{{\rm d}}{{\rm d}s} f_k(s) =\sqrt{2\pi}\,(2\pi s)^2\,e^{-\frac{1}{2}(2 \pi s)^2(1+k)^2}
-\int_0^{2\pi s}  \frac{{\rm d}\omega}{\sqrt{2\pi}} \, 2\pi k(\omega+2\pi k s)\omega^2\,e^{-\frac{1}{2}(\omega+2\pi s k)^2}\ .
\end{equation}
Then, observing that
$$
e^{-x^2}\leq\frac{1}{1+x^2}
$$
and that the second integral is always positive,
\begin{equation}
\frac{{\rm d}}{{\rm d}s} f_k(s)\leq\frac{\sqrt{2\pi}(2\pi s)^2}{1+2\pi^2s^2(k+1)^2}
\leq \frac{2\sqrt{2\pi}}{(1+k)^2}\ .
\end{equation}
On the other hand, we know that
\begin{equation}
 \frac{4\pi}{\sqrt{2\pi}} \sum_{k=0}^\infty \frac{1}{(1+k)^2} =  \frac{4\pi}{\sqrt{2\pi}} \frac{\pi^2}{6} <+\infty,
\end{equation}
therefore, using the Weierstrass M-test (see e.g.,~\cite{Knapp}), we can conclude that the sequence $\{ \frac{{\rm d}}{{\rm d}s} f_k(s) \}_k$
is uniformly convergent in any open interval of the kind $(0,S)$, with $S<\infty$.
It is also clear that the sequence $\{ f_k(s) \}_k$ converges in $s=0$.
Then, as a consequence of Theorem~$1.23$ in~\cite{Knapp},
we have that $\{ f_k(s) \}_k$
is uniformly convergent in any compact interval $[0,S]$.

\section{Analytical approximation of \texorpdfstring{$\Phi(\lambda,q,\omega)$}{}}\label{app.5}

In the following, we provide an approximated expression of  the function $\Phi(\lambda,q,\omega)$ {introduced in~\eqref{eq.46s}} as a linear combination of special functions. As a byproduct, we will also derive another proof of the fact that $\lim_{\lambda\rightarrow +\infty}\alpha(\lambda)=+\infty$. The function  $\Phi(\lambda,q,\omega)\coloneqq \partial_q\Psi(\lambda,q,\omega)$ is given by
\begin{align}\label{eq.106}
&\partial_q\bigg(\sum_{k\in\Z}\int_{\epsilon_1(\lambda,q,\omega,k)}^{\epsilon_2(\lambda,q,\omega,k)}\frac{\de z}{\sqrt{2\pi}}e^{-\frac{z^2}{2}}\bigg)=\sum_{k\in\Z}\exp\bigg(-\frac{(\pi+2k\pi-\lambda\sqrt{q}\omega)^2}{2\lambda^2(1-q)}\bigg)\Bigg(\frac{-\omega\lambda+\sqrt{q}\pi(1+2k)}{2\lambda\sqrt{q}(1-q)^{3/2}}\Bigg)\nonumber\\
&\hspace{6cm}-\exp\bigg(-\frac{(2k\pi-\lambda\sqrt{q}\omega)^2}{2\lambda^2(1-q)}\bigg)\Bigg(\frac{-\omega\lambda+2k\pi\sqrt{q}}{2\lambda\sqrt{q}(1-q)^{3/2}}\Bigg),
\end{align}
where we have used the explicit expression of $\epsilon_1(\lambda,q,\omega,k)$ and $\epsilon_2(\lambda,q,\omega,k)$,
\begin{equation}
\epsilon_1(\lambda,q,\omega,k)\coloneqq \frac{-\lambda\sqrt{q}\omega+2k\pi}{\lambda\sqrt{1-q}},\qquad\epsilon_2(\lambda,q,\omega,k)\coloneqq
\frac{-\lambda\sqrt{q}\omega+\pi+2k\pi}{\lambda\sqrt{1-q}}\ .
\end{equation}
To construct the desired approximation of $\Phi(\lambda,q,\omega)$, we replace the Gaussian terms appearing in~\eqref{eq.106} with Lorentzian functions, namely, we set
\begin{equation}\label{eq.111}
    \exp\bigg(-\frac{(\pi+2k\pi-\lambda\sqrt{q}\omega)^2}{2\lambda^2(1-q)}\bigg)\simeq\frac{1}{1+\frac{(\pi+2k\pi-\lambda\sqrt{q}\omega)^2}{2\lambda^2(1-q)}}=
    \frac{2\lambda^2(1-q)}{2\lambda^2(1-q)+(\pi+2k\pi-\lambda\sqrt{q}\omega)^2},
    \end{equation}
    and
   \begin{equation}\label{eq.112}
    \exp\bigg(-\frac{(2k\pi-\lambda\sqrt{q}\omega)^2}{2\lambda^2(1-q)}\bigg)\simeq\frac{1}{1+\frac{(2k\pi-\lambda\sqrt{q}\omega)^2}{2\lambda^2(1-q)}}=\frac{2\lambda^2(1-q)}{2\lambda^2(1-q)+(2k\pi-\lambda\sqrt{q}\omega)^2}.
    \end{equation}
 Note that, these approximation are the more accurate, the closer is $q$ to $1$ from below, or the greater is $\lambda$. Inserting~\eqref{eq.111} and~\eqref{eq.112} into~\eqref{eq.106} yields a new function $\widetilde{\Phi}(\lambda,q,\omega)$:
\begin{align}\label{eq.110}
  \widetilde{\Phi}(\lambda,q,\omega)=\sum_{k\in\Z}\frac{\lambda}{\sqrt{q}\sqrt{1-q}}&\Bigg(\frac{-\omega\lambda+\sqrt{q}\pi(1+2k)}{2\lambda^2(1-q)+(\pi+2k\pi-\lambda\sqrt{q}\omega)^2}\nonumber\\
  &\hspace{2cm}  -\frac{-\omega\lambda+2k\pi\sqrt{q}}{2\lambda^2(1-q)+(2k\pi-\lambda\sqrt{q}\omega)^2}\Bigg)\ .
  \end{align}
  We now show that the series in~\eqref{eq.110} can be exactly summed, and provides an approximation of the function $\Phi(\lambda,q,\omega)$. Indeed, consider first the term
    \begin{align}\label{eq.114}
        &\frac{\lambda}{\sqrt{q}\sqrt{1-q}}\sum_{k\in\Z}\frac{-\omega\lambda+\sqrt{q}\pi(1+2k)}{2\lambda^2(1-q)+(\pi+2k\pi-\lambda\sqrt{q}\omega)^2}\nonumber\\[0.1cm]
        &=\frac{\lambda}{\sqrt{q}\sqrt{1-q}}\sum_{k\in\Z}\frac{-\omega\lambda+\sqrt{q}\pi(1+2k)}{\Big(\pi+2k\pi-\lambda\sqrt{q}\omega+\sqrt{2\lambda^2(q-1)}\Big)\Big(\pi+2k\pi-\lambda\sqrt{q}\omega-\sqrt{2\lambda^2(q-1)}\Big)}\nonumber\\
        &=\frac{2\lambda\sqrt{q}-\omega\sqrt{2\lambda^2(q-1)}}{8\pi\sqrt{q}\sqrt{1-q}}\sum_{k\in\Z}\frac{1}{k+\bigg(\frac{1}{2}+\frac{\sqrt{2\lambda^2(q-1)}-\lambda\sqrt{q}\omega}{2\pi}\bigg)}\nonumber\\[0.1cm]
    &\hspace{5.2cm}+\frac{\omega\sqrt{2\lambda^2(q-1)}+2\sqrt{q}\lambda}{8\pi\sqrt{q}\sqrt{1-q}}\sum_{k\in\Z}\frac{1}{k+\bigg(\frac{1}{2}-\frac{\lambda\sqrt{q}\omega+\sqrt{2\lambda^2(q-1)}}{2\pi}\bigg)}\ .
    \end{align}
    Let us consider separately the contribution to the series for $k\in[-\infty,-1]$ and $k\in[0,\infty]$. For the negative values of $k$, we have:
    \begin{align}\label{eq.123ne}
    &\frac{2\lambda\sqrt{q}-\omega\sqrt{2\lambda^2(q-1)}}{8\pi\sqrt{q}\sqrt{1-q}}\sum_{k=-\infty}^{-1}\frac{1}{k+\bigg(\frac{1}{2}+\frac{\sqrt{2\lambda^2(q-1)}-\lambda\sqrt{q}\omega}{2\pi}\bigg)}\nonumber\\[0.1cm]
    &\hspace{5.2cm}+\frac{\omega\sqrt{2\lambda^2(q-1)}+2\sqrt{q}\lambda}{8\pi\sqrt{q}\sqrt{1-q}}\sum_{k=-\infty}^{-1}\frac{1}{k+\bigg(\frac{1}{2}-\frac{\lambda\sqrt{q}\omega+\sqrt{2\lambda^2(q-1)}}{2\pi}\bigg)}\nonumber\\[0.1cm]
    &=-\frac{2\lambda\sqrt{q}-\omega\sqrt{2\lambda^2(q-1)}}{8\pi\sqrt{q}\sqrt{1-q}}\sum_{k=1}^{+\infty}\frac{1}{k-\bigg(\frac{1}{2}+\frac{\sqrt{2\lambda^2(q-1)}-\lambda\sqrt{q}\omega}{2\pi}\bigg)}\nonumber\\[0.1cm]
    &\hspace{5cm}-\frac{\omega\sqrt{2\lambda^2(q-1)}+2\sqrt{q}\lambda}{8\pi\sqrt{q}\sqrt{1-q}}\sum_{k=1}^{+\infty}\frac{1}{k-\bigg(\frac{1}{2}-\frac{\lambda\sqrt{q}\omega+\sqrt{2\lambda^2(q-1)}}{2\pi}\bigg)},
    \end{align}
    where we have changed the sign of $k$ to $-k$. It is worth observing that the series in~\eqref{eq.123ne} is divergent, as one expects upon considering that the series~\eqref{eq.114} will converge only considering the values of $k\in [-\infty,+\infty]$. Indeed, sending $k$ into $k+1$, and adding and subtracting the factor $\frac{1}{k+1}$ in~\eqref{eq.123ne}, we get
    \begin{align}\label{eq.120}
    &-\frac{2\lambda\sqrt{q}-\omega\sqrt{2\lambda^2(q-1)}}{8\pi\sqrt{q}\sqrt{1-q}}\sum_{k=0}^{+\infty}\frac{1}{k+\bigg(\frac{1}{2}-\frac{\sqrt{2\lambda^2(q-1)}-\lambda\sqrt{q}\omega}{2\pi}\bigg)}\nonumber\\[0.1cm]
    &\hspace{5.2cm}-\frac{\omega\sqrt{2\lambda^2(q-1)}+2\sqrt{q}\lambda}{8\pi\sqrt{q}\sqrt{1-q}}\sum_{k=0}^{+\infty}\frac{1}{k+\bigg(\frac{1}{2}+\frac{\lambda\sqrt{q}\omega+\sqrt{2\lambda^2(q-1)}}{2\pi}\bigg)}\nonumber\\
    =&\frac{4\sqrt{q}\lambda}{8\pi\sqrt{q}\sqrt{1-q}}\,\gamma-\frac{4\sqrt{q}\lambda}{8\pi\sqrt{q}\sqrt{1-q}}\sum_{k=1}^{+\infty}\frac{1}{k+1}\nonumber\\
    &+\frac{2\sqrt{q}\lambda-\omega\sqrt{2\lambda^2(q-1)}}{8\pi\sqrt{q}\sqrt{1-q}}\psi^{(0)}\bigg(\frac{1}{2}-\frac{\sqrt{2\lambda^2(q-1)}-\lambda\sqrt{q}\omega}{2\pi}\bigg)\nonumber\\[0.1cm]
    &+\frac{\omega\sqrt{2\lambda^2(q-1)}+2\sqrt{q}\lambda}{8\pi\sqrt{q}\sqrt{1-q}}\psi^{(0)}\bigg(\frac{1}{2}+\frac{\lambda\sqrt{q}\omega+\sqrt{2\lambda^2(q-1)}}{2\pi}\bigg),
    \end{align}
    where $\gamma$ is the so-called \textit{Euler-Mascheroni} constant, while $\psi^{(0)}(z)$ is the \textit{digamma function}~\cite{leb1972,olve2014} --- namely, the function defined as
    \begin{equation}
    \psi^{(0)}(z)\coloneqq\frac{\Gamma^\prime(z)}{\Gamma(z)}=-\gamma+\sum_{k=0}^{+\infty}\frac{1}{k+1}-\frac{1}{z+k},\;\; \mathrm{Re}\,z>0.
    \end{equation}
     As  anticipated above, a divergent term appears in~\eqref{eq.120}. Nevertheless, considering the contribution to the series~\eqref{eq.114} for $k\in[0,+\infty]$ yields
    \begin{align}\label{eq.126ne}
    &\frac{2\lambda\sqrt{q}-\omega\sqrt{2\lambda^2(q-1)}}{8\pi\sqrt{q}\sqrt{1-q}}\sum_{k=0}^{+\infty}\frac{1}{k+\bigg(\frac{1}{2}+\frac{\sqrt{2\lambda^2(q-1)}-\lambda\sqrt{q}\omega}{2\pi}\bigg)}\nonumber\\[0.1cm]
    &\hspace{5.2cm}+\frac{\omega\sqrt{2\lambda^2(q-1)}+2\sqrt{q}\lambda}{8\pi\sqrt{q}\sqrt{1-q}}\sum_{k=0}^{+\infty}\frac{1}{k+\bigg(\frac{1}{2}-\frac{\lambda\sqrt{q}\omega+\sqrt{2\lambda^2(q-1)}}{2\pi}\bigg)}\nonumber\\[0.1cm]
     =&-\frac{4\lambda\sqrt{q}}{8\pi\sqrt{q}\sqrt{1-q}}\,\gamma+\frac{4\lambda\sqrt{q}}{8\pi\sqrt{q}\sqrt{1-q}}\sum_{k=0}^{\infty}\frac{1}{k+1}\nonumber\\
     &-\frac{2\lambda\sqrt{q}-\omega\sqrt{2\lambda^2(q-1)}}{8\pi\sqrt{q}\sqrt{1-q}}\psi^{(0)}\bigg(\frac{1}{2}+\frac{-\lambda\sqrt{q}\omega+\sqrt{2\lambda^2(q-1)}}{2\pi}\bigg)\nonumber\\[0.1cm]
    &-\frac{\omega\sqrt{2\lambda^2(q-1)}+2\sqrt{q}\lambda}{8\pi\sqrt{q}\sqrt{1-q}}\psi^{(0)}\bigg(\frac{1}{2}-\frac{\lambda\sqrt{q}\omega+\sqrt{2\lambda^2(q-1)}}{2\pi}\bigg).
    \end{align}
    Combining equations~\eqref{eq.120} and~\eqref{eq.126ne}, we obtain the final expression
\begin{align}\label{eq.114fin}
&\frac{\lambda}{\sqrt{q}\sqrt{1-q}}\sum_{k\in\Z} \frac{-\omega\lambda+\sqrt{q}\pi(1+2k)}{2\lambda^2(1-q)+(\pi+2k\pi-\lambda\sqrt{q}\omega)^2}\nonumber\\
&=C_1(\lambda,q,\omega)\Bigg(\psi^{(0)}\!\bigg(\frac{1}{2}+Z(\lambda,q,\omega)\bigg)-\psi^{(0)}\!\bigg(\frac{1}{2}-Z(\lambda,q,\omega)\bigg)\Bigg)\nonumber\\
&\hspace{4cm}+C_2(\lambda,q,\omega)\Bigg(\psi^{(0)}\!\bigg(\frac{1}{2}+W(\lambda,q,\omega)\bigg)
-\psi^{(0)}\!\bigg(\frac{1}{2}-W(\lambda,q,\omega)\bigg)\bigg),
\end{align}
where we set
    \begin{align}
&Z(\lambda,q,\omega)=\frac{\lambda\sqrt{q}\omega-\sqrt{2\lambda^2(q-1)}}{2\pi},\qquad W(\lambda,q,\omega)=\frac{\lambda\sqrt{q}\omega+\sqrt{2\lambda^2(q-1)}}{2\pi}\nonumber\\
&C_1(\lambda,q,\omega)=\frac{2\sqrt{q}\lambda-\omega\sqrt{2\lambda^2(q-1)}}
       {8\pi\sqrt{q}\sqrt{1-q}},\qquad C_2(\lambda,q,\omega)=\frac{\omega\sqrt{2\lambda^2(q-1)}+2\sqrt{q}\lambda}
       {8\pi\sqrt{q}\sqrt{1-q}},
\end{align}
and where, as expected, the divergent contributions are no longer present. Analogously, we can sum the second term in the series~\eqref{eq.110} thus arriving at the following final expression of the function $\widetilde{\Phi}(\lambda,q,\omega)$:

    \begin{align}
    \widetilde{\Phi}(\lambda,q,\omega)&=C_1(\lambda,q,\omega)\Bigg(\psi^{(0)}
\bigg(\frac{1}{2}+Z(\lambda,q,\omega)\bigg)-\psi^{(0)}
\bigg(1+Z(\lambda,q,\omega)\bigg)
     +\psi^{(0)}\bigg(-Z(\lambda,q,\omega)
     \bigg)\nonumber\\
     &\hspace{-0.6cm}-\psi^{(0)}\bigg(\frac{1}{2}-Z(\lambda,q,\omega)\bigg)
     \Bigg)+
    C_2(\lambda,q,\omega)
    \Bigg(\psi^{(0)}\bigg(\frac{1}{2}+W(\lambda,q,\omega)\bigg)
   -\psi^{(0)}
    \bigg(\frac{1}{2}-W(\lambda,q,\omega)\bigg)\nonumber\\
&\hspace{-0.6cm}+\psi^{(0)}\bigg(-W(\lambda,q,\omega)\bigg)-
\psi^{(0)}\bigg(1+W(\lambda,q,\omega)\bigg)\Bigg).
    \end{align}
A close inspection of the function $\widetilde{\Phi}(\lambda,q,\omega)$ shows that its modulus, $\big|\widetilde{\Phi}(\lambda,q,\omega)\big|$, actually provides an upper bound to $\Phi(\lambda,q,\omega)$, which is consistent with the approximation of Gaussian functions by Lorentzian ones considered in~\eqref{eq.111} and~\eqref{eq.112}.

Using the well known asymptotic expansion of the digamma function~\cite{olve2014}:
\begin{equation}
\psi^{(0)}(z)\sim \ln z-\frac{1}{2z},\quad \abs{z}\rightarrow +\infty,
\end{equation}
we can also derive the asymptotic expansion of the function $\widetilde{\Phi}(\lambda,q,\omega)$, for $\lambda\rightarrow \infty$, i.e.,
\begin{align}\label{eq.59n}
\widetilde{\Phi}(\lambda,q,\omega)&\overset{\lambda \rightarrow \infty}{\simeq} C_1(\lambda,q,\omega)\left(\ln\Bigg(\frac{\pi-\sqrt{2\lambda^2(q-1)}+\lambda\sqrt{q}\omega}{2\pi+\lambda\sqrt{q}\omega-\sqrt{2\lambda^2(q-1)}}\Bigg)\right.\\
\nonumber
&\left.
-\ln\Bigg(\frac{\pi-\lambda\sqrt{q}\omega+\sqrt{2\lambda^2(q-1)}}{-\lambda\sqrt{q}\omega+\sqrt{2\lambda^2(q-1)}}\Bigg)\right)\nonumber\\[0,1cm]
&+C_2(\lambda,q,\omega)\left(\ln\Bigg(\frac{\pi+\lambda\sqrt{q}\omega+\sqrt{2\lambda^2(q-1)}}{\pi-\lambda\sqrt{q}\omega-\sqrt{2\lambda^2(q-1)}}\Bigg)-\ln\Bigg(\frac{2\pi+\lambda\sqrt{q}\omega+\sqrt{2\lambda^2(q-1)}}{-\lambda\sqrt{q}\omega-\sqrt{2\lambda^2(q-1)}}\Bigg)\right),
\end{align}
from which, it is not difficult to show that $\lim_{\lambda\rightarrow \infty}\widetilde{\Phi}(\lambda,q,\omega)=0$. On the other hand, taking into account that $\lim_{\lambda\rightarrow \infty}\Psi(\lambda,q,\omega)=1/2$ (see~\eqref{Psi}), it follows that $\lim_{\lambda\rightarrow \infty}\widetilde{\Phi}(\lambda,q,\omega)/\Psi(\lambda,q,\omega)=0$, from which we can argue, once again, the limit  $\lim_{\lambda\rightarrow \infty}\alpha(\lambda)=+\infty$.

\section*{Acknowledgments}
The authors acknowledge financial support from the PNRR MUR Project PE0000023-NQSTI and from the PRIN MUR Project 2022SW3RPY. GG is partially
supported by Istituto Nazionale di Fisica Nucleare (INFN) through the project ``QUANTUM"
and by the Italian National Group of Mathematical Physics (GNFM-INdAM). GG acknowledges financial support from the University of Bari through the 2023-UNBACLE-0245516
grant.

\end{document}